\documentclass[reqno]{amsart}
\usepackage{amsfonts}
\usepackage{amsthm}
\usepackage{amssymb}
\usepackage{dsfont}
\usepackage{amscd}
\usepackage{color}
\usepackage{graphicx}

\newtheorem{theorem}{Theorem}[section]
\newtheorem{corollary}[theorem]{Corollary}
\newtheorem{lemma}[theorem]{Lemma}
\newtheorem{prop}[theorem]{Proposition}
\theoremstyle{definition}

\theoremstyle{remark}
\newtheorem{remark}[theorem]{Remark}

\newtheorem*{acknow}{Acknowledgments}
\numberwithin{equation}{section}

\def\be{\begin{equation}}
\def\ee{\end{equation}}
\def\ba{\begin{eqnarray*}}
\def\ea{\end{eqnarray*}}
\def\bae{\begin{eqnarray}}
\def\eae{\end{eqnarray}}
\def\bc{\begin{center}}
\def\ec{\end{center}}

\begin{document}

\title{A.~Hurwitz and the origins of random matrix theory in mathematics}

\author{Persi Diaconis} \address{Department of Mathematics and Statistics, Stanford University, 
50 Serra Mall, Bldg. 380, Stanford,
CA 94305-2125}
\author{Peter J. Forrester} \address{Department of Mathematics and Statistics, The University of Melbourne, Victoria 3010, Australia; ARC Centre of Excellence for Mathematical \& Statistical Frontiers
}\email{p.forrester@ms.unimelb.edu.au}


\begin{abstract}
The purpose of this article is to put forward the claim that Hurwitz's paper ``{\" U}ber die Erzeugung der Invarianten durch Integration.'' [G{\" o}tt. Nachrichten (1897), 71-90] should be regarded as the origin of random matrix theory in mathematics. Here Hurwitz introduced and developed the notion of an invariant measure for the matrix groups $SO(N)$ and $U(N)$. 
He also specified a calculus from which the explicit form of these measures could be computed in terms of an appropriate parametrisation --- Hurwitz chose to use Euler angles. This enabled him to define and compute invariant group integrals over $SO(N)$ and $U(N)$. 
{His main result can be interpreted
probabilistically: the Euler angles of a uniformly distributed matrix
are independent with beta distributions (and conversely). We use this
interpretation to give some new probability results.}
How Hurwitz's ideas and methods show themselves in the subsequent work of Weyl, Dyson and others on foundational studies in random matrix theory is detailed.
\end{abstract}
\maketitle

\maketitle
\section{Introduction}
{  Random matrix
theory, as a pure discipline, is the study of matrix theory in the setting that the matrices belong to an ensemble and are thus specified probabilistically.
It is a healthy subject in its own right, with applications in quantum physics, wireless engineering, number theory,
statistics and many other parts
of mathematics and its applications. Textbooks and reviews on the subject include \cite{Me91, De99, Di03, TV04, AGZ09, Fo10, BS10, PS11,ABD11, Ve10,Ta12,Tr15}. }
It is common knowledge in random matrix theory that the origins of the subject in theoretical physics can be traced back to the work of Wigner in the 1950's (see e.g.~\cite{Wi57}; this along with much other source material from the 50's and 60's is reprinted in \cite{Po65}). Wigner introduced large real symmetric random matrices as a model of the statistical properties of the highly excited energy levels of complex nuclei. Significant developments to this theme were made by Dyson in the early 1960's; see e.g.~\cite{Dy62} along with other articles as reprinted in \cite{Po65}. The articles \cite{Bo05} and
\cite{FM09} review how random matrix appeared in nuclear physics from an historical point of view, supplementing
the article by Porter in the introduction of  \cite{Po65}.

Asking now about the origins of random matrix theory in mathematical statistics, common knowledge then singles out the 1928 paper of Wishart on correlation matrices \cite{Wi28}. The exact functional form of the joint eigenvalue probability density function was subsequently established in four independent works by Fisher, Girshick, Hsu and Roy each published in 1939. The review article \cite{An07}, in addition to containing the citation to these works, gives some details of the methods of derivation. It is also revealed in this article that a fifth independent work deriving the joint eigenvalue probability density function published by Mood in 1951 was also discovered in the year 1939. The functional form was reported in the 1943 textbook by Wilks \cite{Wi43}, and it is through this source that Wigner learnt of it for subsequent applications in nuclear physics, as reported in \cite{Wi57}.

In fact mathematical statistics and nuclear physics are often coupled together in statements about the first studies of random matrices. Thus the following quotes from well known text books on random matrix theory: ``Random matrices were first encountered in mathematical statistics by Hsu, Wishart and others in the 1930's, but an intensive study of their properties in connection with nuclear physics began only with the work of Wigner in the 1950's." \cite{Me67, Me91}; 
``Random matrices have been a part of advanced statistical analysis since the end of the 1920s with the work of
Wishart on fixed-sized matrices with Gaussian entries. The first asymptotic results on the limiting spectrum of large
random matrices were obtained by Wigner in the 1950s is a series of papers motivated by nuclear physics" \cite{TV04};
``Indeed, the study of random matrices, and in particular the properties of their eigenvalues, has emerged from the applications, first in data analysis (in the early days of statistical sciences, going back to Wishart), and later as statistical models for heavy nuclei atoms, beginning with the seminal work of Wigner.'' \cite{AGZ09}; ``Initiated in the 1920's -- 1930's by statisticians and introduced in physics in the 1950's --1960's by Wigner and Dyson \ldots'' \cite{PS11}.

Let us now pose the question as to the origins of random matrix theory in mathematics.  Actually very few articles ever address this point in their introductory remarks.
One exception is in the preface of Forrester's 2010 book `Log-gases and random matrices'  \cite{Fo10} where one reads: ``Long before their occurrence in physics, random matrices appeared in mathematics, especially in relation to the Haar measure on classical groups. Perhaps the first work of this type is due to Hurwitz, who computed the volume form of a general unitary matrix parametrized in terms of Euler angles.'' The work of Hurwitz referred to was published in 1897 \cite{Hu97}.

{ It is the purpose of the present article to put forward the case that Hurwitz's work, in addition to having the historical significance of
being the first substantial mathematical work contributing to modern day random matrix theory, contains ideas of lasting importance to the subsequent
development of the field, and as such is deserves being better known and appreciated. In a nutshell, in \cite{Hu97} Hurwitz introduced the notion
of an invariant measure for the classical groups $SO(N)$ and $U(N)$; he showed how upon the introduction of a parametrisation in terms of Euler
angles the measure factorises, and he used the factorisation to compute the implied volume.
As explained below, Hurwitz's
construction has a simple, equivalent 'random matrix' interpretation:
the Euler angles of a uniformly distributed matrix in any of
$SO(N)$, $U(N)$ (and $Sp(2N)$) are independent with explicit beta distributions.
This gives a simple algorithm for generating such matrices with invariant measure,
a novel representation of their eigenvalues and some new theorems
about the traces and moments of matrix coefficients. Some of this relates to contemporary topics in present day research into random matrix theory.
In previous eras Hurwitz's work underpins the works of Weyl, Dyson and others on the computation of the eigenvalue probability density function
for the classical groups, and the specification of the invariant measure for other random matrix ensembles.

We begin in Section two by describing Hurwitz's motivation from invariant theory.
Section three describes Euler angles for $SO(N)$ and $U(N)$. It gives a
version of Hurwitz's argument in modern language, with stand alone proofs. Section four
interprets Hurwitz's results in the language of { contemporary} random matrix theory,
computing volumes and 
{ showing their relevance to the computation of the normalisation for 
the eigenvalue probability density of Dyson's three
circular ensembles. Also given are extensions to $Sp(2N)$ and Mat$(S_N)$, the latter referring to the set of $N \times N$ permutation
matrices.} Section five continues with applications to Monte
Carlo simulations and some probabilistic results that follow directly
from Hurwitz construction.
Here is an example (Corollary \ref{C5.3} below): Let $Y_i$ ($i=1,2,\dots$) be independent random
variables specified by $Y_i = Z_i/ \sqrt{(Z_1^2+...+Z_i^2)}$,  with each $Z_i$ a standard
normal variate. Let $Y = Y_1 Y_2 + Y_2 Y_3 + Y_3 Y_4 + \cdots$.
Then $Y$ has a standard normal distribution.}

\section{Invariant theory}
A little (on-line) research soon reveals that the A.~Hurwitz of \cite{Hu97} was the younger of two mathematician brothers, Julius and Adolf \cite{OS14}. He was a doctoral student of Klein and a teacher of Hilbert.

Prominent during the second half of the 1800's was the study of invariants and covariants of homogeneous polynomials in $n$ variables --- so called $n$-aray forms \cite{Cr86, Pa06}. Let the $n$-aray form be denoted $F(\vec{a}, \vec{x}$), where $\vec{a}$ is the vector of coefficients and $\vec{x}$ the vector of unknowns. Suppose that under the linear change of $\vec{x} = A\vec{x} \,';  \: A \in GL(n)$, one has $F(\vec{a}, A\vec{x} \, ') = F(\vec{a} \, ' , \vec{x} \, ')$.  This can be taken as the definition of the coefficients $\vec{a} \, ' $, which are linear combinations of the coefficients $\vec{a}$. Given $F$ and $A \in GL(n)$ arbitrary, a multi-variable polynomial $C(\vec{a}, \vec{x}$) is said to be covariant with respect to $F$ if 
\begin{equation}\label{C1}
C(\vec{a}\,', \vec{x}\,') = (\det A)^{w}C(\vec{a}, \vec{x})
\end{equation}
for some $w \in \mathbb{Z}^+$. If $C$ is furthermore independent of $\vec{x}$ it is termed an invariant.

As a concrete example, consider the general degree two $n$-array form
$$
F(\vec{a}, \vec{x}) = \sum_{i=1}^{n} a_{ii}x_{i}^{2} + 2\sum_{1 \le i < j \le n}a_{ij}x_{i}x_{j}
$$
Thus the vector of coefficients consists of $n(n+1)/2$ components. Specifying that $a_{ij} = a_{ji}$ these are naturally written as the coefficient matrix $K = [a_{ij}]_{i,j=1,\ldots,n}$. With $\vec{x}$ regarded as a column vector we can then write $F(\vec{a}, \vec{x}) = \vec{x}^{\, T}K\vec{x}$. Writing $K\,'$ = [$a\,'_{ij}]_{i,j=1,\ldots,n}$ we see from the definition that $K\,' = A^{T}KA$. Hence, for any $p \in \mathbb Z^{+}$ (\ref{C1})  is satisfied by the choice  $C(\vec{a}, \vec{x}) = (\det A)^{p}$ and $w=2p$. Being independent of $\vec{x}$, this is an invariant.

Following \cite{Ha86}, Cayley's Finiteness Problem was to determine --- if possible --- a finite number of covariants such that every covariant could be represented as a polynomial in this generating set. The Finiteness problem was solved for binary forms by Gordan in 1868, then by Hilbert for $n$-array forms.

As explained clearly in \cite{Ha86}, Hilbert's work had the significant implication that an affirmative solution of the Finiteness Problem could be deduced from the construction of a linear operator $\mathcal D$ defined on homogeneous polynomials $B$ with the properties
\begin{equation}\label{8}
\mathcal D[B] \in \mathcal J, \quad \mathcal D[P] = P, \quad \mathcal D[BP] = \mathcal D[B]P
\end{equation}
Here $\mathcal{J}$ is a set of homogeneous polynomials which are invariants in some generalised sense, and  $P \in \mathcal J$. Moreover, Hilbert promoted the problem of studying invariants with respect to some subgroup $G$ of $GL(n)$. Hilbert was able to show that his methods extend to the case of $G = SO(3)$, but not $SO(n)$ with $n > 3$. It is the Finiteness Problem for $SO(n)$ that Hurwitz solved in \cite{Hu97} and in so doing, as we will see, initiated random matrix theory in mathematics.

According to \cite{Ha86}, in the case of a finite subgroup $G$ of $ GL(n)$, it had been observed in the 1890's that for $F$ a positive definite Hermitian form the operator 
\begin{equation}\label{9}
J[F] = \frac{1}{|G|} \sum_{S\in G} F(Sx)
\end{equation}
gives an invariant of $G$. Hurwitz observed that for any $F$, (\ref{9})  defines an invariant of $G$, and that all invariants can be written in this way. He furthermore realised that $J$ has the properties (\ref{8}), thus implying an affirmative answer to the Finiteness Problem for finite subgroups of $GL(n)$. Crucial for random matrix theory is his ``fertile idea'' (quoting from \cite{Ha86}): `` [in (\ref{9})] $J$ is defined by summation over the group $G$; for continuous groups such as the rotation group in $n$-space, the analogue of $J$ may be defined by replacing summation over the group by integration. In this way Hurwitz solved the Finiteness Problem for the $n$-space rotation group in 1897.''

Before giving some of the details of Hurwitz's paper \cite{Hu97} as they relate to random matrix theory, let us quote from the encyclopedia.com article on Hurwitz: ``To obtain orthogonal invariants he devised the invariant volume and integration in the orthogonal groups (Werke, paper no.81), which, generalized to compact groups by I.~Schur and H.~Weyl and complemented by the invention of Haar's measure, have become extremely powerful tools in modern mathematics. This is one of the fundamental discoveries for which Hurwitz's name will be remembered [the other being his theorem \cite{Hu98} on real, normed division algebras, proving that the real, complex, quaternion and octonion number systems exhaust all possibilities]."

As this article contends, Hurwitz's work on the invariant volumes and integrations over the orthogonal and unitary groups marks the origin of random matrix theory in mathematics, and thus Hurwitz's name should also be associated with random matrix theory for this historical fact. Remarkably, it is furthermore a fact that Hurwitz's theorem on the classification of real, normed division algebras plays a fundamental role in random matrix theory \cite{Dy62}.

\section{Euler parametrization and invariant measure of the orthogonal and unitary groups}
\subsection{Parametrization of the orthogonal group $SO(N)$}
With $I_{p}$ denoting the $p \times p$ identity matrix, define $R_{j}(\theta) \in SO(N)$, $j=1,\ldots,N-1$, as the block diagonal matrix
\begin{equation}\label{R}
R_{j}(\theta) = \begin{bmatrix} I_{j-1}& \ & \ & \ \\ \ & \cos \theta & \sin \theta & \ \\ \ & -\sin \theta & \cos \theta & \ \\ \ & \ & \ & I_{N-j-1}
\end{bmatrix}
\end{equation}
Hurwitz \cite{Hu97} showed that these matrices could be used to sequentially define matrices $E_{j} \in SO(N)$ $(j = N-1,\dots,1)$ with the property that
\begin{equation}\label{VE}
\begin{bmatrix} V_{j+1} & \ \\ \ & I_{N-j-1}
\end{bmatrix}
= \begin{bmatrix} V_{j} & \ \\ \ & I_{N-j}
\end{bmatrix} E_{j},
\end{equation}
where $V_{l} \in SO(l)$. Here $V_{j+1}$ is to be thought of as given, with the aim being to construct $E_{j}$ so that the structure (\ref{VE}) results. Assuming that this is possible, for a general $V_{N} \in SO(N)$ we will then have the factorization 
\begin{equation}\label{VE1}
V_{N} = E_{1} E_{2} \cdots E_{N-1}.
\end{equation}

\begin{lemma}\label{L1}
For certain angles, referred to as Euler angles,
\begin{equation}\label{tp}
0 \le \theta_{1,j+1} < 2 \pi, \ 0 \le \theta_{i,j+1} \le \pi \ (2 \le i \le j \le N - 1)
\end{equation}
the equation (\ref{VE}) can be satisfied by choosing 
\begin{equation}\label{tp1}
E_{j} = R_{j} (\theta_{j,j+1}) R_{j-1} (\theta_{j-1,j+1}) \cdots R_{1} (\theta_{1,j+1}).
\end{equation}
{ The choice of Euler angles is unique except in cases that $\theta_{i,j+1} = 0$ or $\pi$.}
\end{lemma}

\noindent 
{\it Proof.}
Observe that with $\vec{x} = (x_{1},\ldots,x_{N})$ a row vector we have $\vec{x}  R_{l}^{T} = \vec{x}\,'$ where 
\begin{equation}\label{xcs}
x ' _{l} = cx_{l} - sx_{l+1},\ x ' _{l+1} = sx_{l} + cx_{l+1},\ x '_{i} = x_{i},  \quad (i \neq l) 
\end{equation}
and we have written $c = \cos\theta, s = \sin\theta$. Thus by choosing $\theta$ such that $\tan\theta = x_{l} / x_{l+1}$ if $x_{l+1} \neq 0, \theta = \pi/2$ if $x_{l+1} = 0$, we have $x'_{l} = 0$. In particular, applying (\ref{xcs}) in the case $l=1$ and with respect to row $j+1$ transforms the block $V_{j+1}$ in (\ref{VE}) to have a zero entry in the bottom left corner. We then apply (\ref{xcs}) in the cases $l=2$, $l=3,\ldots, l=j$ in order, all with respect to row $j+1$ to transform the block $V_{j+1}$ in (\ref{VE}) to have all entries zero in the bottom row except the final entry. In the case of $V_{3}$, this process can be illustrated by the following diagram (cf.\cite{Di00})
\begin{equation}\label{tp2}
\begin{bmatrix} 
* & * & * & \ \\
* & * & * & \ \\
* & * & * & \ \\
\ & \ & \ & I_{N-3}
\end{bmatrix} \xrightarrow{R_1^T(\theta_{1,3})}
\begin{bmatrix} 
* & * & * & \ \\
* & * & * & \ \\
0 & * & * & \ \\
\ & \ & \ & I_{N-3}
\end{bmatrix} \xrightarrow{R_2^T(\theta_{2,3})}
\begin{bmatrix} 
* & * & * & \ \\
* & * & * & \ \\
0 & 0 & * & \ \\
\ & \ & \ & I_{N-3}
\end{bmatrix}
\end{equation}

Since all matrices in the reduction are orthogonal the final entry in the bottom row of the reduced form of $V_{j+1}$ must equal 
$\pm 1$; its value can be chosen to be $+1$ by using the freedom of allowing $\theta_{1,j+1} \in [0, 2\pi)$ (recall (\ref{tp})). Moreover, the reduced form of $V_{j+1}$ being an orthogonal matrix also implies all entries in column $j+1$, except the diagonal entry, are zero. Thus, with $E_j$ given by (\ref{tp1}) and for approximate angles in the ranges (\ref{tp}), (\ref{VE}) is established.
\hfill $\square$

\medskip
We remark that the $N=3$ case of (\ref{VE1}) gives for a general $V_3 \in SO(3)$ the decomposition
$$
V_3 = \begin{bmatrix}
\cos\phi & \sin\phi & 0 \\
-\sin\phi & \cos\phi & 0 \\
0 & 0 & 1 \end{bmatrix}
\begin{bmatrix}
1 & 0 & 0 \\
0 & \cos\theta & \sin\theta \\
0 & -\sin\theta & \cos\theta \end{bmatrix}
\begin{bmatrix}
\cos\psi & \sin\psi & 0 \\
-\sin\psi & \cos\psi & 0 \\
0 & 0 & 1 \end{bmatrix},
$$
where $0 \le \theta \le \pi, \quad 0 \le \phi, \psi < 2\pi$. Geometrically this corresponds to rotations about the $z$-axis, the transformed $x$-axis, then the transformed $z$-axis, as first identified by Euler \cite{Eu00}. The latter work, in
which (\ref{VE1}) is also outlined for general $N$, is referenced in \cite{Hu97}.
The number of independent parameters in (\ref{VE1}) is $1+2+\cdots+(N-1) = \frac{1}{2}N(N-1)$. Hurwitz set about computing the volume form in terms of these parameters. This requires the specification of a measure. Since the aim from the invariant theory viewpoint was to make use of a continuum analogue of (\ref{9}), the measure had to be chosen to be invariant under the group action of multiplication by a fixed $V_0 \in SO(N)$.

\subsection{The invariant measure of $SO(N)$}
Consider the half line $\mathbb{R}^+$. For $c>0$ the invariant measure $d\mu(x)$ for the multiplicative group of positive real numbers must have the property
$$
d\mu(cx) = d\mu(x)
$$
and is thus given by $d\mu(x) = dx/x$. In the case of the orthogonal group it was realised by Hurwitz that an invariant measure is obtained by the choice
\begin{equation}\label{mu}
d\mu = (V^TdV),
\end{equation}
{where the notation $(\cdot)$ used on the RHS denotes the product of independent differentials of the matrix of differentials in question.}
To make sense of this, note from the relation $V^TV = I_N$ that $V^TdV$ is anti-symmetric. With $A = [A_{i,j}]_{i,j=1,\dots,N}$ anti-symmetric, the line element $ds$ corresponding to the Euclidean metric is specified by 
\begin{equation}\label{mu0}
(ds)^2 = {\rm Tr}(dA^TdA) = 2\sum_{1\le j<k\le N}{(dA_{j,k})^2},
\end{equation}
which implies the volume form { and thus $d \mu$} is given by
\begin{equation}\label{mu1}
(dA) = 2^{N(N-1)/4}\prod_{1\le j<k \le N}{dA_{jk}}.
\end{equation}
The fact that $d\mu(V_0V) = d\mu(V), V_0 \in SO(N)$ is immediate from (\ref{mu}). To see the invariance of (\ref{mu}) under right multiplication by $V_0$ requires the fact that for $A$ antisymmetric
and $X$ fixed and real,
\begin{equation}\label{13.1}
( X^TdA \, X ) = ( \det X^TX )^{(N-1)/2}(dA)
\end{equation}
(see e.g.~ \cite[Ex.1.3 q.2]{Fo10}). The invariance now follows since with $X=V_0$, $\det X^TX = 1$. 

Hurwitz \cite{Hu98} evaluated (\ref{mu}) in terms of the Euler angles of Lemma \ref{L1}.

\begin{prop}\label{P3.2}
The invariant measure (\ref{mu}) as further defined by (\ref{mu1}) is given in terms of the Euler angles of Lemma 1 by 
\begin{equation}\label{P1}
d\mu = 2^{N(N-1)/4}\prod_{1\le j<k \le N}(\sin{\theta_{j,k}})^{j-1}d\theta_{j,k}.
\end{equation}
\end{prop}

\noindent 
{\it Proof.}
It follows from (\ref{VE}) in the case $j = N-1$ and the product rule for differentiation that 
$$
dV_N =
\begin{bmatrix}
dV_{N-1} &  \ \\
\ & 0
\end{bmatrix} E_{N-1} +
\begin{bmatrix}
V_{N-1} & \ \\
\ & 1
\end{bmatrix} dE_{N-1}.
$$
Consequently
$$
E_{N-1} V_{N}^{\, T} dV_{N} \, E_{N}^{\, T} =
\begin{bmatrix}
V_{N-1}^{\, T} dV_{N-1} & \ \\
\ & 0
\end{bmatrix} + dE_{N-1} E_{N-1}^{\, T}.
$$
After making use of (\ref{13.1}) this latter formula shows that the contribution to $(V_{N}^{\, T}dV_{N})$ from ($dE_{N-1}E_{N-1}^{\, T})$ comes entirely from the final row of the latter, to be denoted $(d\vec{E}_{N-1}^{\ (N)}E_{N-1}^{\, T})$ say, and thus
\begin{equation}\label{E3}
(V_{N}^{\, T} dV_{N}) = 2^{(N-1)/2}(V_{N-1}^{\, T} dV_{N-1})(d\vec{E}_{N-1}^{(N)} E_{N-1}^{\, T})
\end{equation}
(the factor $2^{(N-1)/2}$ is in keeping with the corresponding factor in (\ref{mu1})).

Details of the computation of $(d\vec{E}_{N-1}^{\, (N)} E_{N-1}^{\, T})$ are not given in \cite{Hu97}. We follow the strategy given in \cite{Fo10} for the corresponding quantity in the computation of the invariant measure for $U(N)$. First, according to the calculus of wedge products (see e.g.~\cite[1.2.1]{Fo10}), and with $\{\vec{u}\}_j$ denoting the $j$-th component of the vector $\vec{u}$, we have that up to a possible sign
$$
(d\vec{E}_{N-1}^{\, (N)} E_{N-1}^{\, T}) = \det{\left[ \Big\{ \frac{\partial \vec{E}_{N-1}^{\, T}}{\partial \theta_{j,N}}  E_{N-1}^{\,T} \Big\}_k \right]_{j,k=1,\ldots,N-1}}.
$$
Introducing the further notation $\{A\}_{j,k}$ for the entry $(j,k)$ of the matrix $A$, the determinant can be factorized to read
\begin{equation}\label{14.1}
\det{ \Bigg[ \bigg[ \Big\{ \frac{\partial \vec{E}_{N-1}^{\,T}}{\partial \theta_{j,N}} \Big\}_{k'} \bigg]_{j = 1,\dots,N-1 \atop k' = 1,\dots,N}   \bigg[ \Big\{E_{N-1}^{\,T} \Big\}_{k', k}\bigg]_{k'=1,\dots,N \atop k=1,\dots,N-1} \Bigg]}.
\end{equation}

According to the Cauchy-Binet theorem (see e.g.~\cite[Eq.~(6.88)]{Fo10}) this latter determinant is equal to the sum of the product of determinants of the matrix obtained by blocking out the $l$-th column of the first matrix and the $l$-th row of the second. We see from the definition (\ref{tp1}) that the entry in position $k'$ of $\vec{E}_{N-1}^{\, (N)}$ is equal to
\begin{equation}\label{14.2}
 \prod_{l=k'}^{N-1} \sin{\theta}_{l,N} \ \cos{\theta}_{k'-1,N}
\end{equation}
where $\theta_{0,N} = 0$. This allows all entries in the first matrix of the product (\ref{14.1}) to be computed explicitly. Denoting this matrix with column $(l)$ deleted by a superscript $(l)$, $[\cdot] \rightarrow [\cdot]^{(l)}$, we see that $[\cdot]^{(l)}$ has either all entries below the diagonal equal to zero, or all entries one diagonal below the diagonal zero. In the first case the value of the determinant is immediate, while in the second, row operations can be used to reduce it to diagonal form. In both circumstances we find
$$
\det{\bigg[ \Big\{ \frac{\partial \vec{E}_{N-1}^{\ N}}{\partial \theta_{j,N}} \Big\}_{k'} \bigg]_{j = 1,\dots,N-1 \atop k' = 1,\dots,N}} = (-1)^{N+l-1} \Big\{ E_{N-1} \Big\}_{N,l}  \prod_{j=1}^{N-1} (\sin{\theta_{j,N}})^{j-1}. 
$$

Consequently (\ref{14.1}), after noting $\{E_{N-1}\}_{N,l} = \{E_{N-1}^{\, T}\}_{l,N}$, is equal to
\begin{equation}\label{14.2a}
\prod_{j=1}^{N-1} (\sin{\theta_{j,N}})^{j-1} \sum_{l=1}^{N} (-1)^{N+l-1} \Big\{ E_{N-1}^{\, T} \Big\}_{l,N} \det\bigg[ \Big\{ E_{N-1}^{\, T} \Big\}_{k', k} \bigg]_{k' = 1,\dots,N \atop k =1,\dots,N-1}^{(l)},
\end{equation}
where here $[\cdot]^{(l)}$ denotes that row $l$ has been deleted. We recognise the sum in (\ref{14.2a}) as the Laplace expansion of $\det E_{N-1}^{\, T}$ by the final column. But $E_{N-1}^{\, T} \in SO(N)$ so the summation in (\ref{14.2}) is equal to unity and we thus have
$$
( dE_{N-1}^{\, T} E_{N-1} ) = \prod_{j=1}^{N-1} ( \sin{\theta_{j,N}} )^{j-1}.
$$
Substituting in (\ref{E3}) and iterating gives (\ref{P1}).
\hfill $\square$

\medskip
\begin{remark} The fact that (\ref{P1}) is strictly positive unless $\phi_{i,j+1} = 0$ or $\pi$ for any $2 \le i \le N -1$ when it vanishes, is in keeping
with Euler angle parametrisation being unique except for these cases, as noted in the statement of Lemma \ref{L1}.
\end{remark}
{ \begin{remark}
The factorization in (\ref{P1}) shows that the Euler angles
$\theta_{i,j}$ are independent under the invariant measure. Conversely, as
discussed in Section \ref{S5.2} below,  (\ref{P1}) gives
an algorithm for choosing a random matrix from the invariant measure
by multiplying independent rotations {  with specific distributions} in the $(j,k)$ planes.
\end{remark}}

\subsection{ The unitary group $U(N)$ }

Hurwitz \cite{Hu97} computed the invariant measure for the unitary group using analogous reasoning. The elementary rotation matrices (\ref{R}) depending on a single Euler angle $\theta$ must now be replaced by matrices $U_j (\phi, \psi, \alpha), j = 1,\ldots , N-1$, depending on three Euler angles and specified by
\begin{equation}\label{UQ}
U_j = 
\begin{bmatrix}
I_{j-1} & \ & \ & \ \\
\ & \cos\phi \ e^{i\alpha} & \sin\phi \ e^{i\psi} & \ \\
\ & -\sin\phi \ e^{-i\psi} & \cos\phi \ e^{-i\alpha} & \ \\
\ & \ & \ & I_{N-j-1}.
\end{bmatrix}
\end{equation}
The initial task is to show that for certain Euler angles
\begin{equation}\label{UQa}
0 \leq \phi_{i,j+1} \leq \frac{\pi}{2}, \quad  0 \leq \psi_{i,j+1} < 2 \pi \quad (i \leq j), \quad 0 \leq \alpha_j < 2\pi
\end{equation}
the equation (\ref{VE}) with $V_l \in SU(l)$ can be satisfied by choosing $E_j$ in (\ref{VE1}) according to the formula
\begin{multline}\label{E4}
E_j = U_j ( \phi_{j, j+1}, \psi_{j,j+1}, 0)  U_{j-1} ( \phi_{j-1, j+1}, \psi_{j-1, j+1}, 0 ) \\ \times \cdots \times
U_1(\phi_{1,j+1}, \phi_{1,j+1}, \alpha_{j+1}).
\end{multline}
{ Explicit construction (see e.g.~\cite{Mu62}) shows that this parametrisation is unique except for $\phi_{i,j+1} = 0$ or $\pi/2$.}
Consequently $V_N \in U(N)$ can be decomposed
\begin{equation}\label{E5}
V_N = e^{i\alpha_1} E_{1} E_{2} \cdots E_{N-1}.
\end{equation}
The mechanism underlying (\ref{E4}) is the unitary analogue of that illustrated in (\ref{tp2}).

Minor modification of the reasoning which tells us that the choice (\ref{mu}) is an invariant measure for the orthogonal group shows that
\begin{equation}\label{mu5}
d\mu = ( V^\dagger dV )
\end{equation}
is the corresponding invariant measure for the unitary group. For the meaning of the RHS, note that $V^\dagger dV$ is 
anti-Hermitian, and thus equal to $i$ times an Hermitian matrix. The analogue of (\ref{mu0}) and (\ref{mu1})  is\footnote{Hurwitz gives a further factor of $\sqrt{N!}$ on the RHS.}  
$$
(ds)^2 = {\rm Tr} (dH^\dagger dH) = \sum_{j=1}^N (dH_{j j})^2 + 2 \sum_{1 \leq j < k \leq N} (dH_{j k}^{(r)})^2 + (dH_{j k}^{(i)})^2
$$
\begin{equation}\label{mu6}
( dH ) = 2^{N((N-1)/2} \prod_{j=1}^N dH_{j j} \prod_{1 \leq j < k \leq N} dH_{j k}^{(r)} dH_{jk}^{(i)}
\end{equation}
where $(r), (i)$ refer to the real and imaginary parts respectively. Note that there are $N + N(N-1) = N^2$ independent parameters in (\ref{mu6}), which is the same number as in (\ref{E5}): $1 + 3 + \cdots + (2N - 1) = N^2$.

Hurwitz \cite{Hu97} gave the explicit form of (\ref{mu5}) in terms of the Euler angles of (\ref{E4}). His derivation was based on the analogue of (\ref{E3}), but as with (\ref{P1}) no details were given; a proof along the lines of that given above for
Proposition \ref{P3.2} can be found in \cite[\S 3.2.1]{Fo10}.

\begin{prop}\label{P3.3}
The invariant measure (\ref{mu5}) as further defined by (\ref{mu6}) is given in terms of the Euler angles of (\ref{E4}) by
\begin{equation}\label{17.1}
d\mu = 2^{N(N-1)/2} \prod_{1 \leq j < k \leq N } \cos \phi_{j,k} ( \sin \phi_{j,k} )^{2j-1} d\phi_{j,k}  d\psi_{j,k} \prod_{l=1}^N d\alpha_l.
\end{equation}
\end{prop}

{ \begin{remark} Analogous to (\ref{P1}), the decomposition (\ref{17.1}) shows that under the invariant measure the Euler angles
$\phi_{j,k}, \psi_{j,k}$ and the phases $\alpha_l$ are all independent. A construction of the corresponding marginal distributions
is given in Section \ref{S5.2} below. \end{remark}}

Thus we see that Hurwitz gave both a specification of the invariant measure for the orthogonal and unitary groups in terms
of a volume form involving the matrix elements, and in terms of an Euler angle parametrisation. In the final section of
\cite{Hu97}, the former construction is extended to arbitrary Lie groups diffeomorphic to $\mathbb R^n$.

 In the early
1930's Haar  proved the existence of an invariant measure --- now called the Haar measure --- on any separable
compact topological group, and soon after von Neumann proved uniqueness.
{ The standard  treatment of Haar measure is  \cite{Ha76}
with \cite{HR63} giving many explicit examples (and
counter-examples). The recent book  \cite{DS14} 
has many further references.}

\section{Significance in random matrix theory}
\subsection{Volumes}

The volume of $SO(N)$, vol$\big( SO(N) \big)$ say, is defined as the integral of (\ref{P1}) over the allowed range of Euler angles (\ref{tp}). As noted by Hurwitz \cite{Hu97}, performing this calculation gives
\begin{equation}\label{W1a}
{\rm vol} \big( SO(N) \big) = \frac{1}{2} \ 2^{N(N+3)/4} \prod_{k = 1}^N \frac{\pi^{k/2}}{\Gamma(k/2)}.
\end{equation}
Without the factor of $1/2$, this gives vol$\big( O(N) \big)$. A set of matrices related to $O(N)$ is $O(N)/ \big( O(1) \big)^N$. This is realized by requiring that the first entry in each column be positive. This reduces the volume of $O(N)$ by $2^{-N}$, so we have
\begin{equation}\label{W2}
{\rm vol} \Big( O(N)/ \big( O(1) \big)^N \Big) = 2^{N(N-1)/4} \prod_{k=1}^N \frac{\pi^{k/2}}{\Gamma(k/2)}.
\end{equation}

Although not made explicit in \cite{Hu97}, knowledge of (\ref{17.1}) allows us to similarly compute
\begin{equation}\label{W3}
{\rm vol} \big( ( U (N) \big) = 2^{N(N+1)/2} \prod_{k=1}^N \frac{\pi^k}{\Gamma(k)}.
\end{equation}
The set of matrices $U(N) / \big( U(1) \big)^N$ is realized by requiring that the first component of each column be real positive reducing the volume of $U(N)$ by $(2\pi)^{-N}$, and thus
\begin{equation}\label{2P}
{\rm vol} \Big( U(N) / \big( U (1) \big)^N \Big) = 2^{N(N-1)/2} \prod_{l=1}^{N-1} \frac{\pi^l}{\Gamma(l+1)}.
\end{equation}

One of the first applications of knowledge of these volumes came in relation to Dyson's decomposition \cite{Dy62a} of the invariant measure, $d\mu(S)$ say, for the space of symmetric unitary matrices $\{U^TU\}$ where $U \in U (N)$ is chosen with the corresponding invariant measure (\ref{mu5}). It was subsequently noted in \cite{Dy70} that this space is isomorphic to the symmetric space $U (N) / O (N)$. Taking into consideration that the infinitesimal real symmetric matrix $dM$ relates to $dH := (iU)^\dagger dU$ by $dM_{ij} = 2 dH_{ij}^{(r)}$ \cite[Eq.~(111)]{Dy62a} we have \cite[Eq.~(ii4)]{Dy62a}
\begin{equation}\label{19.1}
d\mu(S)d\mu(R) = 2^{\frac{1}{2}N(N+1)}d\mu(U),
\end{equation}
where $S \in U(N) / O(N), \ R \in O(N), \ U \in U(N)$.
Consequently
\begin{equation}\label{2U}
{\rm vol} \big( U(N) / O(N) \big) = 2^{N(N+1)/2} {\rm vol} \big( U(N) \big) / {\rm vol} \big(O(N) \big) = 2^{3N/4} \prod_{l=1}^N \frac{\pi^{(l+1)/2}}{\Gamma \big((l+1)/2\big)},
\end{equation}
where use has been made of (\ref{W1a}), (\ref{W3}) and the duplication formula for the gamma function.

For $S \in U(N) / O(N)$, Dyson introduced the  diagonalization $S = Q^TLQ$ \ where \ $Q \in O(N)/\big( O(1) \big)^N$ is the matrix of eigenvectors and $L = {\rm diag} (e^{i\theta_1}, \ldots, e^{i\theta_N})$ the matrix of the (ordered) eigenvalues. He furthermore decomposed the corresponding invariant measure
\begin{equation}\label{20.1}
d\mu(S) = \prod_{1 \leq j < k \leq N}| e^{i\theta_k} - e^{i\theta_j} | \, d\theta_l \cdots d\theta_N\ d\mu(Q).
\end{equation}
Relaxing the requirement that the eigenvalues be ordered, this implies that the eigenvalue probability density function is given by
$$
\frac{1}{C_N} \prod_{1 \leq j < k \leq N} | e^{i\theta_k} - e^{i\theta_j} |, \quad 0 \leq \theta_l \leq 2\pi \ ( l = 1, \ldots, N ),
$$
where
\begin{equation}\label{CN}
C_N = \int_0^{2\pi} d\theta_1 \cdots \int_0^{2\pi} d\theta_N  \prod_{1 \leq j < k \leq N} | e^{i\theta_k} - e^{i\theta_j} |.
\end{equation}
Significantly, knowledge of the volumes (\ref{W2}) and (\ref{2U}) implies the value of the normalization (\ref{CN}),
$$
C_N = N! \frac{{\rm vol} \big(U(N) / O(N) \big)}{{\rm vol} \Big( O(N) / \big(O(1)\big)^N \Big)}
$$
as seen from (\ref{20.1}). This shows \cite[eq.(130)]{Dy62}
$$
C_N = \frac{\Gamma (N/2 + 1)}{\big( \Gamma (3/2) \big)^N}.
$$

More can be said in relation to (\ref{W1a}) and (\ref{W3}). Define the unit $(n-1)$-sphere by the equation $x_1^2 + \cdots + x_n^2 = 1$. Its surface area is equal to $A_{n-1} = 2\pi^{n/2} / \Gamma(n/2)$, while the radius $R$ $(n-1)$-sphere has surface area $A_{n-1}(R) = R^{n-1}A_{n-1}(1)$. Thus
\begin{equation}\label{21.1}
\frac{{\rm vol}\big( SO(N) \big)}{{\rm vol}\big( SO(N-1) \big)} = A_{N-1} (\sqrt{2}), \quad \frac{{\rm vol}\big( U(N) \big)}{{\rm vol}\big( U(N-1) \big)} = {1 \over \sqrt{2}}A_{2N-1}(\sqrt{2}).
\end{equation}

The former of these is a corollary of (\ref{E3}), and the latter a corollary of the unitary analogue of (\ref{E3}), supplemented by the fact that $\int d\vec{E}^{\ (N)}_{N-1} E^{\ T}_{N-1} = A_{2N-1}(1)$ in the unitary case. These facts can in turn be interpreted geometrically as saying $\vec{E}_{N-1}^{(N)}$ is distributed uniformly on the surface of the real and complex unit-$(N-1)$- spheres respectively. Dyson \cite{Dy62a} observed this directly, and thus didn't have cause to refer to Hurwitz \cite{Hu97} for the computation of the volumes. 

\subsection{ Decomposed invariant measures }

In his 1939 book on the classical groups Weyl \cite{We39} decomposed the invariant measures for $SO(N)$ and $U(N)$ as identified by Hurwitz in terms of variables corresponding to the eigenvalues and eigenvectors. It is fair to comment that the reference to \cite{Hu97} in \cite{We39} is somewhat oblique. The details of the calculation are, from a technical viewpoint, simpler than the working needed to deduce Propositions \ref{P3.2} or \ref{P3.3}. For definiteness, we will consider $U(N)$.

\begin{prop}
Let $V \in U(N)$ and introduce the eigenvalue decomposition $V = U^\dagger LU$ where $U \in U(N)/\big(U(1)\big)^N$ is the matrix of eigenvectors and $L = {\rm diag} (e^{i\theta_1}, \ldots, e^{i\theta_N})$ the matrix of ordered eigenvalues. One has
\begin{equation}\label{W1}
d\mu(V) = \prod_{1 \leq j < k \leq N} | e^{i\theta_k} - e^{i\theta_j} |^2 \, d\theta_1 \cdots d\theta_N\ d\mu(U).
\end{equation}
\end{prop}

\noindent 
{\it Proof.}
Denote $dM_V = V^\dagger dV$ and $dM_U = U^\dagger dU$. From the definitions
$$
U^\dagger dM_VU = dM_UL - L dM_U + iL [d\theta],
$$
where $ [d\theta ] = {\rm diag} (d\theta_1, \ldots, d\theta_N)$. Using the fact that {\rm Tr}\big(($dM_UL - LdM_U) L [ d\theta ] \big) = 0$, it follows from this and (\ref{mu0}) that
$$
{\rm Tr}(dM_V dM_V^\dagger) = \sum^N_{j, k=1 \atop j \neq k} | (dM_U)_{j,k} (e^{i\theta_j} - e^{i\theta_k}) |^2 + \sum_{j=1}^N (d\theta_j)^2.
$$
Now using (\ref{mu1}), (\ref{W1}) follows.
\hfill $\square$

\medskip
As with (\ref{20.1}), since the eigenvalue and eigenvector portions factorize, one reads off from (\ref{W1}) that the eigenvalue probability function for matrices chosen with Haar measure from $U(N)$ is equal to 
$$
\frac{1}{\tilde{C}_N} \prod_{1 \le j < k \le N} | e^{i\theta_k} - e^{i\theta_j} |^2, \quad 0 \le \theta_l \le 2\pi \ (l =1,\ldots,N),
$$
where
$$
\tilde{C}_N = \int_0^{2\pi} d\theta_1 \cdots \int_0^{2\pi} d\theta_N  \prod_{1 \le j < k \le N} | e^{i\theta_k} - e^{i\theta_j} |^2.
$$
In stating this result, the ordering of the
eigenvalues implicit in (\ref{W1}) has been relaxed. Taking this into consideration, and using the fact that
$$
{\rm vol} \big( U(N) \big) = (2\pi)^N {\rm vol} \Big( U(N) / \big( U(1) \big)^N \Big)
$$
it follows \cite{Dy62a}
$$
\tilde{C}_N = (2\pi)^N N!
$$

While Weyl gives only passing reference to Hurwitz in his development of the invariant measure on the classical groups, Hurwitz \cite{Hu97} is cited prominently by Siegel \cite{Si45} in his construction of an invariant measure on the matrix group $SL_N(\mathbb R)$. Subsequently a decomposition analogous to (\ref{20.1}) was given in Jack and Macbeath \cite{JM59}, and a formula for the volume computed. This involves a factor vol$\big( O(N) \big)$; for its value reference was made to \cite{Hu97}.

\subsection{Unitary symplectic and permutation matrices}
In \cite{We39} Weyl introduced the unitary symplectic group $Sp(2N)$  of $N \times N$ unitary matrices in which each
element is a $2 \times 2$ matrix representation of a real quaternion, and thus of the form
\begin{equation}\label{Q}
\begin{bmatrix} z & w \\ - \bar{w} & \bar{z} \end{bmatrix}.
\end{equation}
This, together with $U(N)$ and $O(N)$ makes up the three classical, compact, continuous groups.

The modulus of (\ref{Q}) is defined as $|z|^2 + |w|^2$. We thus see that a general real quaternion of unit models can be parametrised as the $2 \times 2$ block
in (\ref{UQ}) --- an element of $SU(2)$ --- with angles in the range (\ref{UQa}). Let $\mathbf q$ and $\mathbf Q$ be two real quaternions with unit modulus,
and define the $2 \times 2$ matrix of real quaternions
\begin{equation}\label{B1}
\begin{bmatrix} \mathbf q  \cos \rho  & \mathbf Q  \sin \rho \\ -  \bar{\mathbf Q} \sin \rho & \bar{\mathbf q} \cos \rho \end{bmatrix},
\end{equation}
with $0 \le \rho \le \pi/2$ and where in the representation (\ref{Q}) $ \bar{\mathbf q} $ refers to the Hermitian conjugate. { Note that as
a complex matrix (\ref{B1}) is $4 \times 4$.}

Let $U_j^{\rm q} = U_j^{\rm q}(\rho,\mathbf q, \mathbf Q)$ denote the block matrix in (\ref{UQ}) with each entry 1 replaced by the $2 \times 2$ identity,
and the $2 \times 2$ block replaced by (\ref{B1}). We then have that each $V_N \in Sp(2N)$ can be decomposed as in (\ref{E5}) with 
$e^{i \alpha_1}$ replaced by $\mathbf q_1$ and
$U_l$ in (\ref{E4})
replaced by $U_l^{\rm q}(\rho_{l,j+1}, \mathbf q_{l,j+1}, \mathbf Q_{l,j+1})$, where $ \mathbf q_{l,j+1} = \mathbb I_2$ for $l \ne 1$. 
Set
$ \mathbf q_{l,j+1} =   \mathbf q_{j+1}$. We then have
$ \mathbf Q_{l,j+1} =  \mathbf Q_{l,j+1}(\phi_{l,j+1}, \psi_{l,j+1},\alpha_{l,j+1})$, while $ \mathbf q_j =  \mathbf q_{j}(\phi_{j}, \psi_{j},\alpha_{j})$.
{ The representation is unique provided $\rho_{l,j+1}$ is not equal to 0 or $\pi/2$, and $\phi_{l,j+1}, \phi_{j}$ are not equal to $0, \pi/2$.}
A calculation analogous to that required to derive the volume form (\ref{17.1}) in terms of Euler angles for the invariant measure in the case of $U(N)$ gives the
corresponding result for matrices from $Sp(2N)$.

\begin{prop}\label{P4}
Let $V \in Sp(2N)$. Define the invariant measure by  (\ref{mu5}), noting that $i V^\dagger dV$ is then a self dual quaternion Hermitian matrix 
(see e.g.~\cite[Eq.~(2.6)]{Fo10}) so the analogue of (\ref{mu6}) now has proportionality $2^{N(N-1)}$ and four real differentials in the
product over $j < k$.
In terms of the Euler angles as implied by (\ref{B1}) and the subsequent text, we have
\begin{multline}\label{17.1a}
d\mu = 2^{N(N-1)} \prod_{1 \leq j < k \leq N }  \cos^3 \rho_{j,k} (\sin \rho_{j,k})^{4j - 1} {1 \over 2} \sin 2 \phi_{j,k} \, d \phi_{j,k} d\psi_{j,k}  d \alpha_{j,k} \\
\times \prod_{j=1}^N  {1 \over 2} \sin 2 \phi_j d \phi_j d \psi_j d \alpha_j.
\end{multline}
\end{prop}

{ This expression for the invariant measure shares with (\ref{P1}) and (\ref{17.1}) the feature that all the parameterising angles are distributed
independently.}

We now turn our attention to permutation matrices. The set of all $N \times N$ 0-1 matrices with a single 1 in each row and column and no two 1's in the same row or column --- Mat$(S_N)$ say --- are in bijection with the set of permutations on $N$ symbols $S_N$, which with the operation of composition is itself
a classical group, albeit discrete rather than continuous. These matrices are all real orthogonal, and thus form a subgroup of $O(N)$. The invariant measure
in this case is just the uniform measure, in which each element is chosen with equal probability $1/N!$. In an unpublished manuscript dated the year 2000,
Diaconis and Mallows \cite{DM00} have shown that a random element of Mat$(S_N)$, or equivalently of $S_N$, can be generated by
a factorised product of elementary permutation matrices in direct analogy to Hurwitz's factorisation (\ref{VE1}) as used to specify an invariant measure on
$SO(N)$. 

\begin{prop}\label{P4.3}
Let
\begin{equation}\label{TT}
T_j(\mu) = \begin{bmatrix} I_{j-1} & & & \\
 & 1-\mu & \mu & \\
  & \mu & 1 - \mu & \\
   & & & I_{N-j-1} \end{bmatrix},
   \end{equation}
where $\mu = 1$ with probability $j/(j+1)$ and $\mu = 0$ otherwise. Define
\begin{equation}\label{ET}   
E_j^{\rm p} = T_j(\mu_{j,j+1}) T_{j-1}(\mu_{j-1,j+1}) \cdots T_1(\mu_{1,j+1}).
\end{equation}
We have that 
\begin{equation}\label{ETa}   
E_1^{\rm p} E_2^{\rm p} \cdots E_{N-1}^{\rm p}
\end{equation}
is a uniformly chosen permutation matrix.
\end{prop}

\noindent
{\it Proof. } In the language of permutation matrices, we can understand (\ref{ET}) in terms of the factorisation (\ref{VE}) and the procedure (\ref{tp2}).
For example, with $V_{j+1} \in {\rm Mat}(S_{j+1})$ and chosen uniformly at random, the required action of (\ref{TT}) with top block
$T_1^T(\mu_{1,j+1})$ is as the identity with probability that the final entry in the first column of $V_{j+1}$ is zero, which equals $1/(j+1)$.
The required action is as the matrix form of the transposition $(1,2)$ if the final entry is unity, an event which occurs with probability $j/(j+1)$.

An alternative viewpoint is to consider the action of (\ref{ETa}) on letters in the permutation, for which we interpret $T_j(0)$ as the identity
mapping, and $T_j(1)$ as the transposition $(i,i+1)$. We then see that $E_{N-1}^{\rm p}$ acting on the letter $N$ maps it to $N$ with
probability that $T_{N-1}(\mu)$ is the identity, which is $1/N$; it maps $N$ to $N-1$ with probability that $T_{N-1}(\mu) = (N-1,N)$ and
$T_{N-2}(\mu)$ is the identity, which is ${N-1 \over N} \cdot {1 \over N-1} = {1 \over N}$; more generally it maps $N$ to $k$ with probability
that $T_l(\mu) = (l,l+1)$, $(l=k,\dots,N-1)$, and that $T_{k-1}(\mu)$ is the identity, which is $(\prod_{l=k}^{N-1} {l \over l+1} ) {1 \over k} = {1 \over N}$.
Hence $E_{N-1}^{\rm p}$ send $N$ uniformly to $\{1,\dots,N\}$. On the other hand, $E_j^{\rm p}$ for $j=1,\dots,N-2$ does not act on the letter $N$,
so $N$ is mapped uniformly at random by (\ref{ETa}). We now repeat this argument, starting with the action of $E_{N-1}^{\rm p}$ on $N-1$, then the action of
$E_{N-2}^{\rm p}$ on $N-2$ etc., showing that $E_j^{\rm p}$ sends $j+1$ uniformly at random to $\{1,\dots,j+1\}$, to conclude that all letters are
mapped to random positions uniformly. \hfill $\square$

\begin{remark}
For a general permutation, (\ref{ETa}) can be interpreted as a naive bubble-sort. Given a list of $N$ numbers, one begins at the start
and compares successive pairs. If the pair is out of order it is switched, and if not it is left alone, resulting in the largest element being moved to the end.
The procedure is then repeated until all elements are correctly ordered.
\end{remark}

{  \begin{remark}
As in the case of the invariant measure for the continuous classical groups in terms of Euler angles, the parameters $\mu_{j,k}$ implicit in the factorisation
(\ref{ETa}) are all independently distributed under the uniform distribution on Mat$(S_N)$. \end{remark}}

\subsection{Dyson's circular ensembles}
As recalled in the Introduction, Wigner introduced random matrices into theoretical physics to model the Hamiltonian of a complex quantum system. Generally the Hamiltonian of a quantum system can be represented by an infinite matrix. In Wigner's model the infinite matrix is approximated by a finite matrix, which is required to reproduce the statistical properties of the highly excited states. It is postulated that the latter are determined by a global time reversal symmetry $T$ of the Hamiltonian, or the absence of this symmetry. It is known that such an operator either has the property $T^2 = 1$ or $T^2 = -1$, with the latter only applying to finite dimensional systems with an odd number of spin $1/2$ particles (see e.g.~\cite[\S~1.1.1]{Fo10}). In the case that $T^2 = 1$ --- for example, $T$ corresponding to complex conjugation --- the elements of the matrix can always be chosen to be real. Being a statistical theory, an ensemble of real symmetric matrices is thus sought. As detailed in the introduction to \cite{Po65}, Wigner first specified an ensemble by requiring the independent elements be independently distributed from a probability distribution with mean zero and finite variance. He then considered a particular example in which the diagonal entries are standard Gaussians, while entries above the diagonal a Gaussians with mean zero, variance $1/2$. The probability density on the space of real symmetric matrices is then proportional to $e^{-{\rm Tr}H^2}$. The corresponding measure then has the appealing physical feature of being invariant under conjugation by real orthogonal matrices, and in fact this property, together with the requirement that the entries be independent with mean zero uniquely characterizes this measure.

On the other hand, the requirement that the independent entries be independently distributed is not based on physical principles. To obtain a theory of random matrices in quantum mechanics based entirely on the principle of global time reversal symmetry, Dyson \cite{Dy62} replaced random Hamiltonians in favour of random unitary operators which is to be modelled by finite random unitary matrices. In the absence of time reversal symmetry, the measure on unitary matrices is required to be invariant under left and right multiplication by a fixed unitary matrix, and this is given by the invariant measure introduced by Hurwitz \cite{Hu97}, although Dyson references \cite[p.~188]{We39}.

Dyson \cite{Dy62a} showed that a time reversal symmetry $T^2 = 1$ implies that the unitary matrices must be symmetric. It is straightforward to show that symmetric unitary matrices are of the form $S = U^TU$ for $U \in U(N)$. Moreover, Dyson showed that
\begin{equation}\label{25.1}
d\mu(S) = (U^T)^\dagger dS \, U^\dagger
\end{equation}
defines an invariant measure on this space (cf.~(\ref{mu5})) and that furthermore this, like (\ref{mu5}), is unique up to normalization.
Further, it was shown in \cite{Dy62a} that for $T^2 = -1$ the unitary matrices must be self dual quaternion and thus of the form $\tilde{S} = U^D U$,
where $U^D = Z_{2N}^{-1} U^T Z_{2N}$ with $Z_{2N} = \mathbb I_{2N} \otimes \begin{bmatrix} 0 & -1 \\ 1 & 0 \end{bmatrix}$ and $U \in U(2N)$. The corresponding
invariant measure was identified as $d \mu(\tilde{S}) = (U^D)^\dagger d S \, U^\dagger$.

\section{ Averages and Monte Carlo sampling}
{ This section gives several direct applications of
Hurwitz's representation to random matrix theory. In \S 5.1 we show how it
allows computation of moments and mixed moments of matrix entries. Section 5.2 shows
how to use it to generate uniform random matrices on the computer.
Section 5.3 shows that pieces of Hurwitz's representation give random
matrices with the same eigenvalue distribution, and that this in turn leads to a class of unitary Hessenberg matrices.}

\subsection{Averages of matrix elements}
Suppose we wanted to compute 
the average with respect to the invariant measure
$\langle |R_{j,k}|^{2p} \rangle_{R \in SO(N)}$. The invariance implies the average is independent of the particular entry $(j, k)$, so we are free to choose $(j, k) = (N, N)$. Recalling (\ref{14.2}) and (\ref{VE}) with $j = N-1$, we see that $R_{N,N} = \cos \theta_{N-1,N}$. On the other hand (\ref{P1}) gives that the factor dependent on $\theta_{N-1,N}$ in the expression for the invariant measure in terms of the Euler angles is $(\sin \theta_{N-1,N})^{N-2}$. Thus we have
\begin{equation}\label{23E}
\Big \langle |R_{N,N}|^{2p} \Big \rangle_{R \in SO(N)} = \frac{1}{Z_N} \int_0^\pi |\cos \theta|^{2p} (\sin \theta)^{N-2} d\theta,
\end{equation}
where $Z_N$ is such that the RHS equals unity for $p=0$. Noting that this integral is a trigonometric form of the Euler beta integral (see e.g.~\cite[Ex.~4.1 q.1(i)]{Fo10}) this gives
\begin{equation}\label{23F}
\Big \langle |R_{N,N}|^{2p} \Big \rangle_{R \in SO(N)} = \frac{ \Gamma(p + 1/2)\Gamma(N/2)}{\Gamma(1/2)\Gamma(p + N/2)}.
\end{equation}

An alternative way to derive (\ref{23F}) is to use the fact that any one column (or row) of a member of $SO(N)$ with invariant measure is equal in distribution to a vector $\vec{X}_N$ of $N$ independent standard Gaussians normalized to unity; see the next subsection for an expanded discussion on this point.  The square of one entry is then equal in distribution to that of $a/(a+b)$, where $a \mathop{=}\limits^{\rm d} |\vec{X}_1|^2$, $b \mathop{=}\limits^{\rm d} |\vec{X}_{N-1}|^{2}$. This is equal to ${\rm Beta}[ 1/2, (N-1)/2 ]$, and computing the $p$-th moment gives (\ref{23F}). This viewpoint similarly offers the most efficient way to compute the joint moments along any single row or column, which are thus equal to moments of a certain Dirichlet distribution.

Let us now turn our attention to the computation of the joint moments for elements in distinct rows and columns. Orthogonal invariance tells us that it suffices to consider $\langle |R_{N,N}|^{2p}\,|R_{N-1, N-1}|^{2q} \rangle_{R \in SO(N)}$. Recalling (\ref{P1}), and using (\ref{VE1}) to compute $R_{N-1, N-1}$ we thus have
\begin{multline}\label{KP}
\Big \langle |R_{N,N}|^{2p}\,|R_{N-1, N-1}|^{2q} \Big \rangle_{R \in SO(N)} \\
= \frac{1}{\tilde{Z}_N} \int_0^\pi d\theta_{N-3, N-1} (\sin\theta_{N-3, N-1})^{N-4} \int_0^\pi d\theta_{N-2, N-1} (\sin\theta_{N-2, N-1})^{N-3} \\ \times
\int_0^\pi d\theta_{N-2, N} (\sin\theta_{N-2, N})^{N-3} \int_0^\pi d\theta_{N-1, N} (\sin\theta_{N-1, N})^{N-2} \\ \times
|\cos\theta_{N-1, N}|^{2p} |\cos\theta_{N-2, N-1} \cos\theta_{N-2,N} \cos\theta_{N-1,N} \\
- \cos\theta_{N-3, N-1} \sin\theta_{N-2, N-1} \sin\theta_{N-2, N}|^{2q},
\end{multline}
where $\tilde Z_N$ is such that the RHS equals unity for $p = q = 0$.

To make further progress, we suppose temporarily that $2q \in \mathbb  Z_{\ge 0}$. This allows the final factor in (\ref{KP}) to be expanded using the binomial theorem. Defining
\begin{equation}\label{KP1}
T (\alpha, \beta) = \int_0^\pi |\sin\theta|^\alpha |\cos\theta|^\beta d\theta
\end{equation}
this shows
\begin{multline}\label{KP2}
\Big \langle |R_{N, N}|^{2p}\,|R_{N-1, N-1}|^{2q} \Big \rangle_{R \in SO(N)} \\
= \frac{1}{\tilde Z_N} \sum_{l=0}^q \binom{2q}{2l} T \Big( N-4, 2(q-l) \Big) \Big( T\big( N-3 + 2(q-l), 2l \big) \Big)^2 \\
\times T (N-2, 2p+2l).
\end{multline}
After substituting the evaluation of (\ref{KP1}) in (\ref{KP2}) the resulting summation is seen to be proportional to
$$
\frac{\Gamma (p+1/2) \Gamma(q+1/2) \Gamma(q+N/2-1)}{\Big( \Gamma\big( (N-1)/2+q \big) \Big)^2 \Gamma(N/2+p)} \
{}_3 F_2 \bigg( {\displaystyle 1/2, 1/2+p, -q \atop  \displaystyle N/2+p, 2-N/2-q} \bigg| 1 \bigg)
$$
Furthermore, the ${}_3 F_2$ is balanced, and so can be summed using the Pfaff--Saalsch{\" u}tz identity (see e.g.~\cite{AAR99}). Simplification then gives
\begin{multline}\label{KP3}
\Big \langle |R_{N, N}|^{2p}\,|R_{N-1, N-1}|^{2q} \Big \rangle_{R \in SO(N)} \\
= \frac{\Gamma (p+1/2) \Gamma(q+1/2) \Gamma\big( (N-1)/2+p+q \big) \Gamma(N/2) \Gamma\big( (N-1)/2 \big)}
{\big( \Gamma(1/2) \big)^2 \Gamma\big( (N-1)/2+p \big) \Gamma\big( (N-1)/2+q \big) \Gamma(N/2+p+q)}.
\end{multline}
This agrees with the evaluations given in \cite{Br06, BCS11} obtained without direct use of Euler angles; the latter reference uses instead the
formulas of the type first given by Weingarten \cite{We78}.

{ \begin{remark} Perhaps the first paper addressing the computation of mixed moments of matrix elements over the orthogonal
group is due to James \cite{Ja55}. Given our Hurwitz theme, it is of interest to quote his opening paragraph: ``Since Hurwitz \cite{Hu97} first
introduced the idea of integrating over the orthogonal group, integration with respect to invariant measures on groups has become a
powerful method for the analysis of group representations and the study of invariants. Most work in this direction has been along abstract lines,
however, and very few integrals over groups have actually been evaluated."\end{remark}}

\subsection{Generating matrices uniformly from the invariant measure}\label{S5.2}
For complicated integrands, one approach to numerically estimate matrix integrals over $SO(N)$ or $U(N)$ is via Monte Carlo sampling. In its simplest form this requires generating members of $SO(N)$ or $U(N)$ chosen uniformly with respect to the invariant measure. In fact it was for this very task, in the case of $U(N)$, that Zyczkowski and Kus \cite{ZK94} brought to light Hurwitz's paper \cite{Hu97} in the contemporary random matrix literature. Consider first the case of $U(N)$. According to (\ref{17.1}), uniform sampling requires that $\psi_{j,k}$ and $\alpha_j$ be chosen uniformly on the interval $(0, 2\pi)$. Furthermore, upon noting that $\cos\phi_{j,k} (\sin\phi_{j,k})^{2 j - 1} d\phi_{j,k} \propto d(\sin\phi_{j,k})^{2j }$, we see that the variable $\xi_{j,k}$ related to $\phi_{j,k}$ by $\phi_{j,k} = \arcsin{\xi_{j,k}^{1/2j}}$, must be chosen with uniform distribution in $(0, 1)$. For matrices from $SO(N)$, the form (\ref{P1}) says that for $2 \le j \le k-1$ we should change variables
\begin{equation}\label{AS5a}
\tilde{\phi}_{j,k}(\theta_{j,k}) = \int_0^{\theta_{j,k}} (\sin \theta)^{j-1} \, d \theta
\end{equation}
and sample $\tilde{\phi}_{j,k}$ with uniform distribution in the interval $[0,L]$, where
$$
L = \int_0^\pi (\sin \theta)^{j-1} \, d\theta = {\Gamma(1/2) \Gamma(j/2) \over \Gamma((j+1)/2)} .
$$

Making literal use of (\ref{AS5a}) is not practical due to the need to determine $\theta_{j,k}$ given $\phi_{j,k}$. This can be avoided and a practical prescription
specified by observing from (\ref{AS5a}) that the random variable $\tilde{\phi}_{j,k}(\cos \theta_{j,k})$ has a density function on $(-1,1)$ proportional to
$(1 - s^2)^{(j-2)/2}$. This is recognised as the density function involving a particular combination of independent standard real Gaussians $g_i$, telling us that
\begin{equation}\label{AS5b}
\cos \theta_{j,k} \mathop{=}\limits^{\rm d} {g_{j+1} \over \sqrt{g_1^2 + \cdots + g_{j+1}^2}}.
\end{equation}
An equivalent prescription for the uniform generation of orthogonal matrices with uniform measure has been given in
\cite{AOU87}, but without realising that its theoretical underpinning is due to Hurwitz.

A more direct way of deducing (\ref{AS5b}) is possible. Let $G^{\rm r}$ be a random real $N \times N$ matrix of independent standard Gaussian
entries.  Apply the Gram-Schmidt algorithm to the columns of $G^{\rm r}$, so obtaining the factorisation $G^{\rm r} = QR$, where $Q$ is orthogonal and
$R$ is an upper triangular matrix with all diagonal entries positive. Thus
\begin{equation}\label{QG}
Q = (G^{{\rm r} \, T} G^{\rm r})^{-1/2} G^{\rm r}.
\end{equation}
Since the distribution of $G^{\rm r}$ is unchanged by multiplication  on the left or right by a fixed real orthogonal matrix $Q_0$,
we see that the random real orthogonal matrix (\ref{QG}) also has this property, and us thus distributed uniformly with respect to the invariant 
measure on $O(N)$.
We read off from (\ref{QG}) that any one column of $Q$ is distributed as a normalised standard Gaussian vector
$\vec{g} = (g_1\dots,g_N)/|\vec{g}|$ say. Consider in particular the final column. In Hurwitz's parametrisation this is given by (\ref{14.2}).
Equating to the normalised Gaussian vector gives (\ref{AS5b}) in the case $k=N$, $j=1,\dots,N-1$. We remark that
the construction analogous  to (\ref{QG}) of a uniformly distributed matrix $Z \in U(N)$ is
\begin{equation}\label{QGa}
Z = (G^{\rm c  \dagger} G^{\rm c})^{-1/2} G^{\rm c},
\end{equation}
where $G^{\rm c}$ is a random complex $N \times N$ matrix of independent standard complex Gaussian
entries. 

The coset factorization (\ref{VE}) suggests an alternative method to obtain uniform samples with respect to the invariant measure \cite{DS87}. 
Consider for definiteness the case of $U(N)$.
With $j = N-1$, the matrix $E_{N-1}$ in (\ref{VE}) can be interpreted as an element of the quotient space $U(N)/U(N-1)$. Geometrically, this corresponds to the point on the complex $(N-1)$-sphere determining the axis of the lower-dimensional complex rotations specified by $U(N-1)$ (recall the discussion below (\ref{21.1})). Thus $E_{N-1}$ is the matrix which maps the standard unit vector $\vec{e}_N$ to a randomly chosen point $\vec{z}$ on the complex $(N-1)$-sphere. In keeping with (\ref{QG}), such a vector can be formed out of a column vector of $N$ independent standard complex Gaussians, normalized to unity. Writing the final component of $\vec z$ in polar form $z_N = |z_N| e^{i\theta_N}, 0 \le \theta_N \le 2\pi$, and defining $\vec v = \vec z + e^{i\theta_N}\vec e_N, \vec w = \vec v / |\vec v|$, one can check that
\begin{equation}\label{A5a}
E_{N-1} = -e^{i\theta_N} (I_N - 2\vec w \vec w^{\,\dagger})
\end{equation}
has the sought property $E_{N-1}\vec e_N = \vec z$. The matrices $E_{N-j}$ for $j = 2,\ldots,N-1$ are then constructed in an analogous way making use of a vector of $N-j+1$ independent standard complex Gaussians, and have the block structure
\begin{equation}\label{A56}
E_{N-j} =
\begin{bmatrix}
I_{j-1} & \ \\
\ & -e^{-i\theta_{N+1-j}} (I_{N+1-j} - 2\vec w_{N+1-j} \vec w^{\,\dagger}_{N+1-j})
\end{bmatrix};
\end{equation}
see also the discussions in \cite{Ja05}, \cite{Me07} and \cite{BHNY08}.
The analogous construction for matrices from $O(N)$ is given in \cite{St80}.

\subsection{Sampling from the eigenvalue PDF}
Having constructed matrices chosen uniformly at random with respect to the invariant measure, diagonalisation allows
the corresponding eigenvalue PDF to be sampled. However, if this last task is all that is required, more efficient approaches
are possible, as in fact follows by examination of Hurwitz's factorisation (\ref{VE1}), and it's properties in the setting of the
invariant measure.

First we note from (\ref{VE}) in the case $j=N-1$ that
$$
V_N = \begin{bmatrix} V_{N-1} & \\
&1 \end{bmatrix} E_{N-1}.
$$
Moreover, the matrix $\begin{bmatrix} V_{N-1}^T & \\
&1 \end{bmatrix}$ is independent of $V_N$ with respect to right multiplication (but not left).
Since $V_N$ has been chosen with invariant measure, its distribution is unchanged, and we have
$$
V_N \mathop{=}\limits^{\rm d} \begin{bmatrix} V_{N-1} & \\
&1 \end{bmatrix} E_{N-1}  \begin{bmatrix} V_{N-1}^T & \\
&1 \end{bmatrix}
$$
The matrices $V_N$ and $E_{N-1}$ are thus similar in distribution, and so share the same eigenvalue distribution. This is furthermore true of
arbitrary rearrangements of the factors making up $E_{N-1}$.

\begin{prop}\label{Pd}
The random real orthogonal matrix $E_{N-1}$ defined according to (\ref{tp1}), and with angles specified according to 
(\ref{AS5b}), has the same eigenvalue distribution as matrices chosen uniformly with respect to the invariant measure from $SO(N)$.
This property remains true of the random matrix formed by multiplying the factors $R_{N-1}, R_{N-2}, \dots, R_1$ in the definition of 
$E_{N-1}$ in any order.
\end{prop}

\noindent 
{\it Proof.} It remains to show that the eigenvalue distribution is invariant under changing the order of the factors making up $E_{N-1}$.
These factors have the property that $R_i R_j = R_j R_i$ for $|i - j| \ge 2$. Also, being orthogonal, they have the cyclic property that
$R_i X$ and $X R_i$, where $X$ is a general $N \times N$ matrix, are similar and thus have the same eigenvalues. These two properties together allow any order of
the factors to be reached, starting with $R_{N-1} R_{N-2} \cdots R_1$, without altering the eigenvalues. \hfill $\square$

\medskip
{ One immediate consequence is  an application in probability theory.

\begin{corollary}\label{C5.3}
Let $Z_i$ be independent standard Gaussian random variables.
Let $Y_i = Z_i/ \sqrt{(Z_1^2+ \cdots +Z_i^2)}$, and let 
$Y = Y_1 Y_2 + Y_2 Y_3 + Y_3 Y_4 + \cdots$.
Then Y has a standard normal distribution.
\end{corollary}

\noindent
{\it Proof.}  In Proposition \ref{Pd}, the $R_i$ are independent with simple
functions of $Y_i$ as entries. By direct inspection, the trace of $E_N$
is equal to $Y_1Y_2 + Y_2Y_3 + \cdots +Y_{N-1}Y_N +Y_N$. However, this
trace, the sum of the eigenvalues, is exactly distributed as the trace
of a uniformly distributed orthogonal random matrix. In \cite{DS87} 
(see \cite{Me08} and references therein for further developments) it
is shown that the limit distribution of the trace is standard normal.
The result follows. \hfill $\square$}

\begin{remark} The results of Proposition \ref{Pd} and Corollary \ref{C5.3}, together with their analogue for matrices chosen with invariant measure from $U(N)$, $Sp(2N)$ and $S_N$,
were given in the unpublished manuscript \cite{DM00}. In particular, in relation to $S_N$, consideration of Proposition \ref{P4.3} 
leads to the following analogue of Corollary \ref{C5.3}:
Let $Y_i$ be 1 or 0 with probability $1/i$, $1-1/i$ respectively. Then $Y =
Y_1Y_2 + Y_2Y_3 + ...$ has a Poisson(1) distribution. This result can be traced back
to \cite{Ig81}.
\end{remark}

The matrix $E_{N-1}$ as implied by the definition (\ref{tp1}) is a lower Hessenberg matrix --- all elements above the first upper sub-diagonal are zero.
A property of all unitary, and in particular real orthogonal, Hessenberg matrices is that all nonzero entries can be expressed in terms of the
diagonal entries (see e.g.~\cite[Prop.~2.8.1]{Fo10}). Specifically, for $i=1,\dots,N-1$ define $c_i = \cos \theta_{i,N}$,
$\alpha_{i-1} = (-1)^{i-1} c_i$ and $\rho_{i-1} = (1 - \alpha_{i-1}^2)^{1/2}$, and also set $\alpha_{-1} = -1$ and $\alpha_{N-1} = (-1)^{N-1}$.
Multiplying out the matrices in (\ref{tp1}) shows that the diagonal entries of $E_{N-1}$ are given by $-\alpha_{i-2} \alpha_{i-1}$, $(i=1,\dots,N)$,
the leading upper sub-diagonal entries by $\rho_{i-1}$, $(i=1,\dots,N-1)$, and the entries below the diagonal by
$-\alpha_{j-1} \alpha_i \prod_{l=j}^{i-1} \rho_l$, $(i > j)$. Recalling the distribution of $\cos \theta_{i,N}$ as specified in the text
above (\ref{AS5b}), this same Hessenberg orthogonal matrix (after taking its transpose and restricting $N$ to be even) was identified as having
identical eigenvalue distribution as matrices from $SO(N)$ chosen with invariant measure in \cite{KN04}. The working of 
\cite{KN04} made use of Householder reflection matrices, rather than elementary rotation matrices. Moreover, it was noted in \cite{KN04} that
the Hessenberg structure implies that the sequence of characteristic polynomials $\chi_k(\lambda) = \det (\lambda \mathbb I_k - (E_{N-1})_k)$,
where $(E_{N-1})_k$ denotes the top $k \times k$ sub-block of $E_{N-1}$, satisfies the coupled recurrences
\begin{align}\label{5.13}
\chi_k(\lambda) & = \lambda \chi_{k-1}(\lambda) - \alpha_{k-1} \tilde{\chi}_{k-1}(\lambda) \nonumber\\
\tilde{\chi}_k(\lambda) & = \tilde{\chi}_{k-1}(\lambda) -  \lambda \alpha_{k-1} {\chi}_{k-1}(\lambda) 
\end{align}
for $k=1,\dots,N$, where $\chi_0(\lambda) = \tilde{\chi}_0(\lambda) = 1$ and $\tilde{\chi}_k(\lambda) := \lambda^k \chi_k(1/\lambda)$.
The analogue of (\ref{5.13}) for random matrices from $U(N)$ chosen with invariant measure has been shown in \cite{KS09}
to give rise to a characterisation of the statistical state formed by the eigenvalues in terms of a stochastic differential equation.

The  fact that the eigenvalue distribution of $R_{N-1} R_{N-2} \cdots R_1$ 
is unchanged by reordering \cite{DM00} also has an interesting consequence.
Let $R_1 R_3 \cdots R_{2\lfloor N/2 \rfloor - 1} =: R_{\rm odd}$ and $R_2 R_4 \cdots R_{2\lfloor (N+1)/2 \rfloor - 2} =:
R_{\rm even}$. Both are tridiagonal matrices. Their product $R_{\rm odd} R_{\rm even}$ is thus five-diagonal --- up to signs and with $N \mapsto 2N$
it is precisely the so-called CMV matrix (after Cantero, Moral and Vel\'azque \cite{CMV03}) identified in \cite{KN04,KS09} as generating the
eigenvalues of $SO(2N)$ matrices chosen with invariant measure.

\section{Concluding remarks}

In \cite{Hu97} Hurwitz introduced and developed the notion of an invariant measure for the matrix groups $SO(N)$ and $U(N)$. Specifically the invariant measures were defined as the volume forms $(R^TdR)$ and $(U^\dagger dU)$ respectively. Upon being given a particular parametrization --- for this Hurwitz used Euler angles --- it was shown how the use of length elements for Euclidean metrics corresponding to real anti-symmetric and complex anti-Hermitian matrices, respectively, allowed for the explicit determination of these measures. Armed with an invariant measure Hurwitz was able to define, and provide a calculus for, the computation of group integrals over $SO(N)$ and $U(N)$. As an application, in the case of $SO(N)$ this was used to compute the explicit value of the volume of the group manifold.

It is widely appreciated that Hurwitz's paper \cite{Hu97} was pioneering and highly influential with regards to subsequent works of Schur and Weyl on invariance theory for the compact groups, and similarly with regards to the work of Haar, von Neumann, Weil, Cartan and others on the existence and uniqueness of invariant measures for all locally compact topological groups (see e.g.~\cite{Tk01}).
The discoveries in \cite{Hu97} relating to $SO(N)$ and $U(N)$ are also of lasting importance to random matrix theory, and represent notions and techniques which are fundamental to the subject. Given this, it seems only fair that Hurwitz's name should be associated with the origin of random matrix theory in mathematics. 

\begin{acknow}
The work of PD was supported by NSF grant DMS-1208775
and the Australian Research Council Grant DP130100674, and the work of PJF was supported by the Australian Research Council Centre of Excellence ACEMS, 
and  the Australian Research Council  grant DP140102613. 
Thanks are due to  Sherry Wang for help in preparation of the
manuscript, John Stillwell for a careful reading, Colin Mallows for his help over the years
and to Arun Ram for arranging our meeting at Melbourne.
\end{acknow}


\providecommand{\bysame}{\leavevmode\hbox to3em{\hrulefill}\thinspace}
\providecommand{\MR}{\relax\ifhmode\unskip\space\fi MR }
\providecommand{\MRhref}[2]{%
  \href{http://www.ams.org/mathscinet-getitem?mr=#1}{#2}
}
\providecommand{\href}[2]{#2}

\end{document}